\documentclass[reprint,aps,amsmath,amssymb,twoside,prd,showkeys,superscriptaddress]{revtex4-1}
\pdfoutput=1
\usepackage{verbatim}
\usepackage{comment}
\usepackage{graphicx}
\usepackage[T1]{fontenc}
\usepackage{graphicx}
\usepackage{epsfig}
\usepackage{bm}
\usepackage{tensor}
\usepackage{amsfonts}
\usepackage{amssymb}
\usepackage{float}
\usepackage{amsmath}
\usepackage{subfigure}
\usepackage{dcolumn}
\usepackage{cancel}
\usepackage[dvipsnames]{xcolor}
\usepackage[colorlinks]{hyperref}
\usepackage[utf8]{inputenc}

\hypersetup{
     breaklinks=true,
    pdfstartview={FitH},    
    colorlinks=true,       
    linkcolor=blue,          
    citecolor=red,        
    filecolor=magenta,      
    urlcolor=blue,           
    anchorcolor=green,      
    linktocpage=true
}

\def\doi{http://doi.org}

\newcommand{\be}{\begin{equation}}
\newcommand{\ee}{\end{equation}}

\newcommand{\HCd}{\mathcal{H}}

\def\HCdt0{\tilde{\HCd}_{0}}

\newcommand{\affcam}{DAMTP, Centre for Mathematical Sciences, University of Cambridge, Wilberforce Road, Cambridge CB3 0WA, United Kingdom}
\newcommand{\affcamast}{Kavli Institute of Cosmology (KICC), University of Cambridge, Madingley Road, Cambridge, CB3 0HA, UK}

\newcommand{\affbulg}{Institute for Nuclear Research and Nuclear Energy, Bulgarian Academy of Sciences, Sofia, Bulgaria}
\newcommand{\affmaltaa}{Institute of Space Sciences and Astronomy, University of Malta, Malta, MSD 2080}
\newcommand{\affmaltab}{Department of Physics, University of Malta, Malta, MSD 2080}

\begin{document}
\title{Strengthening extended Gravity constraints with combined systems:\\ \texorpdfstring{$f(R)$}{} bounds from Cosmology and the Galactic Center}

\author{D. Benisty}
\email{db888@cam.ac.uk}
\affiliation{\affcam}\affiliation{\affcamast}
\author{J. Mifsud}
\email{jurgen.mifsud@um.edu.mt}
\affiliation{\affmaltaa}\affiliation{\affmaltab}
\author{J. Levi Said}
\email{jackson.said@um.edu.mt}
\affiliation{\affmaltaa}\affiliation{\affmaltab}
\author{D. Staicova} 
\email{dstaicova@inrne.bas.bg}
\affiliation{\affbulg}
\begin{abstract}
Extended gravity is widely constrained in different astrophysical and astronomical systems. Since these different systems are based on different scales it is not trivial to get a combined constraint that is based on different phenomenology. Here, for the first time (to the best of our knowledge), we combine constraints for $f(R)$ gravity from late time Cosmology and the orbital motion of the stars around the galactic center. $f(R)$ gravity models give different potentials that are tested directly in the galactic center. The cosmological data set includes the type Ia supernova and baryon acoustic oscillations. For the galactic star center data set we use the published orbital measurements of the S2 star. The constraints on the universal parameter $\beta$ from the combined system give: $\beta_{HS}=0.154 \pm 0.109$ for the Hu-Sawicki model, while $\beta_{St}= 0.309 \pm 0.19 $ for the Starobinsky dark energy model. These results improve on the cosmological results we obtain. The results show that {{\it combined constraint}} from different systems yields a stronger constraint for different theories under consideration. Future measurements from the galactic center and from cosmology will give better constraints on models with $f(R)$ gravity.
\end{abstract}

\maketitle


\section{Introduction}

Cosmological measurements from the last few decades show that the general theory of Relativity (GR) is not the complete solution for gravity theories. The measurements from the Type Ia supernova \cite{Pan-STARRS1:2017jku}, Baryon Acoustic Oscillations (BAO) \cite{Addison:2013haa,Aubourg:2014yra,Cuesta:2014asa,Cuceu:2019for} and the Cosmic Microwave Background (CMB) \cite{Planck:2018vyg} give strong evidence at least for one modification beyond GR, which is the Cosmological Constant $\Lambda$ \cite{Perlmutter:1998np,Weinberg:1988cp,Lombriser:2019jia,Copeland:2006wr,Frieman:2008sn,Riess:2019cxk}. However, the question is whether GR$\,+\,\Lambda$ is the fundamental theory of gravity or a small part of a bigger theory remains an open question. Modified theories of gravity are theoretically and observationally appealing, and given the ever increasing sensitivity and precision in upcoming surveys, which provides the exciting possibility of robustly testing them against observational data. Among these alternatives, extensions of GR, like $f(R)$ gravity, can be considered a straightforward and natural approach to retain positive results of Einstein's theory and eventually to extend it at infrared and ultraviolet scales \cite{Capozziello:2011et}. 

Besides cosmological systems, there are new constraints on alternative theories of gravity from the strong gravity regime from the galactic center. The relativistic, compact object in the galactic centre is called $Sgr A^{*}$. The stars orbiting $Sgr A^{*}$ are called the S-stars \cite{Yu:2016nzn,Abuter:2018drb,Do:2019txf,Abuter:2020dou,Amorim:2019hwp,Dialektopoulos:2018iph,Borka:2021omc,Capozziello:2014rva,Capozziello:2014mea,Borka-Jovanovic:2019mya}. A large fraction of these stars have orbits with high eccentricities which puts them under the effect of both extremes of gravity. Thus, they reach high velocities at the pericenter and can be used for constraining modified gravity
\cite{Will:2018ont,Will:1997bb,Scharre:2001hn,Moffat:2005si,Zhao:2005zq,Bailey:2006fd,Deng:2009tg,Barausse:2012da,Borka:2012tj,Enqvist:2013tsa,Borka:2013dba,Capozziello:2014rva,Berti:2015itd,Borka:2015vqa,Zakharov:2016lzv,Zhang:2017srh,Dirkes:2017ecu,Pittordis:2017byg,Hou:2017cjy,Nakamura:2018yaw,Banik:2018ydl,Dialektopoulos:2018iph,Kalita:2018ubo,Will:2018ont,Banik:2019zme,Pittordis:2019kxq,Nunes:2019bjq,Anderson:2019eay,Gainutdinov:2020bbv,Bahamonde:2020bbc,Banerjee:2020rrd,Ruggiero:2020yoq,Okcu:2021oke,deMartino:2021daj,DellaMonica:2021xcf,DAddio:2021xsu}. This offers an ideal test bed on which to examine the strong field behavior of modified gravity theories.

The late time cosmic acceleration may be due to exotic matter components in the Universe. Another possibility is that this cosmic speed-up might be caused within GR by the dark energy. The acceleration could be due to purely gravitational effects that emerge from $f(R)$ gravity models. It may also be the case that GR produces the observational consequences of dark energy through additions in the matter section beyond a cosmological constant \cite{CANTATA:2021ktz}. Alternatively, it may be that the dynamics of dark energy is a result of the additional dynamical behavior of modified theories of gravity, and in particular $f(R)$ gravity models. The aim of our work is to combine data from these different systems and to show that more robust constraints of modified theories of gravity can be obtained from combined systems. Although these systems are from different scales, the combined constraint yields a better bound on the parameters. Combining both the cosmological scale tests and the strong field tests of astrophysical systems allows us to obtain higher precision constraints on the free parameters of different $f(R)$ gravity systems.

In Ref.~\cite{Baker:2014zba} different astrophysical systems have been compared, focusing on their gravitational potential vs. their curvature (Kretschmann scalar). Despite the fact that the two systems we compare and statistically add have different curvatures ($10^{34}\,cm^{-2} $ for S-stars orbits and $10^{-50}\, cm^{-2} $ for late time cosmology) we constrain a common parameter of the theory that should give the same value for different curvatures.

The structure of the paper is the following: Section \ref{sec:the} formulates the theory and the equation of motion. Section \ref{sec:mot} derives the approximate equations for the Hubble rate and for the modified orbits. Section \ref{sec:data} describes the dataset. Section \ref{sec:res} discusses the results. Finally, section \ref{sec:sum} summarizes the results and future prospects.

\section{The Theory} \label{sec:the}

A natural approach to $f(R)$ gravity theories is to replace the Ricci scalar $R$ in the Einstein-Hilbert action with an arbitrary function of the Ricci scalar \cite{delaCruz-Dombriz:2006kob,Sotiriou:2008rp,Mohsenzadeh:2012ka,Nojiri:2017ncd,Nojiri:2010wj,Pogosian:2007sw,Nesseris:2015fqa}
\be 
    S=\frac{1}{16\pi G}\int d^{4}x \sqrt{-g}\,f(R) + S_{m}\,, \label{f_R_action} 
\ee
where $G$ is the Newtonian constant, and $S_{m}$ is the matter term in the action, and $g$ is the determinant of the metric. Any $f(R)$ gravity model should fit the conventional standard cosmology as well as explain the current cosmic acceleration issue and the growing cosmological tensions crisis. Moreover, in order to be able to drive the late time cosmic acceleration, the effective dark energy should be $w_{de} = -1$ asymptotically. The cosmological dynamics in $f(R)$ gravity is analyzed in many works~\cite{Clifton:2011jh,CANTATA:2021ktz} across the various flavors of $f(R)$ gravity. The equivalent of the Einstein equation in $f(R)$ gravity reads
\be 
    f'(R) R_{\mu\nu}-\frac{1}{2}f(R) g_{\mu\nu} - \Box_{\mu\nu} f'(R) = 8\pi G T_{\mu\nu}\,, \label{gravi_eq_fR} 
\ee
where $\Box$ is the usual notation for the covariant D'Alembert operator $\Box\equiv\nabla_{\alpha}\nabla^{\alpha}$ where we interpret the effect of different $f(R)$ gravity models as an extra stress-energy contribution which is possible at background level, and the operator $\Box_{\mu\nu} = \left(\nabla_{\mu}\nabla_{\nu}-g_{\mu\nu} \Box\right)$. Compared to GR, $f(R)$ gravity has one extra scalar degree of freedom, $f'(R)$. The dynamics of this degree of freedom is determined by the trace of the field equations in Eq.~(\ref{gravi_eq_fR}), which gives
\begin{equation}
    \Box f'=\frac{2f-f'R}{3} + \frac{8\pi G}{3}T\,, \label{trace_eq1}   
\end{equation} 
where $T$ is the trace of the stress-energy tensor $T_{\mu\nu}$. It is possible to reduce the $f(R)$ equation of motion Eq.~(\ref{gravi_eq_fR}) to
\be 
    G_{\mu \nu}=8\pi G \left( T_{\mu \nu} + T_{\mu \nu} ^{(\text{eff})} \right)\,, \label{field_eq4} 
\ee
where
\begin{align}
    8\pi GT_{\mu \nu} ^{(\text{eff})}=&\frac{f-f'R}{2}g_{\mu\nu} - \Box_{\mu\nu} f' + (1-f')G_{\mu\nu}\,. \label{tilde_T4}
\end{align}
It is possible to see that when $f(R)-f'(R) \approx \text{const}$, the $f(R)$ equations of motion reduce into GR. The known Hu-Sawicki (HS) model \cite{Hu:2007nk} is a good example of $f(R)$ gravity that remains interesting cosmologically and continues to satisfy astrophysical tests. The action of this model is
\begin{equation}
    f(R)=R-m^2 \frac{c_1 (R/m^2)^n}{1+c_2 (R/m^2)^n}\,, \label{Hu}
\end{equation}
where $c_1$, $c_2$ are two free parameters, $m^2\simeq \Omega_{m0}H^{2}_{0}$ is of the order of the Ricci scalar $R_{0}$, $H_{0}$ is the Hubble constant, $\Omega_{m0}$ is the dimensionless matter density today; and $m$ and $n$ are positive constants with $n$ usually taking positive integer values i.e., $n=1, 2, \cdots$. In the rest of our paper, we assume $n=1$ for simplicity. Ref.~\cite{Basilakos:2013nfa} shows that after simple algebraic manipulations Eq.~(\ref{Hu}) can also be written as
\begin{equation}
    f(R) =  R- \frac{2\Lambda }{1+\left(\beta_{HS} \Lambda /R\right)^n}\,, \label{Hu1}
\end{equation}
where $\Lambda = m^2 c_1/2c_2$ and $\beta_{HS} = 2 c_2^{1-1/n}/c_1$. In this form, it is clear that this model can be arbitrarily close to $\Lambda$CDM, depending on the parameters $\beta_{HS}$ and $n$. Moreover, for $n>0$ it has the limits \cite{Basilakos:2013nfa}
\begin{equation}\label{eq:HS_limits}
    \lim_{\beta_{HS}\rightarrow0}f(R) = R-2\Lambda\,, \quad \lim_{\beta_{HS}\rightarrow \infty}f(R) = R\,.
\end{equation}
Since the HS model tends to $\Lambda$CDM for $\beta_{HS}\rightarrow 0$, it can be considered as a small perturbation around the $\Lambda$CDM model. Therefore, it should come as no surprise that the HS model can successfully pass the solar system tests. 

The Starobinsky dark energy model~\cite{Starobinsky:2007hu}, henceforth referred to as the Starobinsky model, is also an interesting model of this class of $f(R)$ gravity in that it has an impact both in inflation as well as at later times of cosmic evolution. This model is defined as
\begin{equation} \label{Star}
    f(R)=R-c_1~m^2 \left[1-\left(1+R^2/m^{4}\right)^{-n}\right]\;.
\end{equation}
where $c_1, m, n$ are free parameters. This can equivalently be represented as
\begin{equation} \label{stargr}
    f(R)=R- 2\Lambda \left(1+\left(\frac{R}{\beta_{St} \Lambda }\right)^2\right)^{-n}
\end{equation}
where $\Lambda= c_1 m^2/2$ and $\beta_{St} = 2/c_1$, and where we obtain identical limits as in Eq.~\eqref{eq:HS_limits} for the $\beta_{St}$ parameter. Indeed, this representation is advantageous because the limit to standard cosmology is more clearly seen.

\section{Solution for different systems} \label{sec:mot}

In this section, we describe the phenomenological predictions for both the cosmological and astrophysical scales of observations under consideration. We do this to ultimately combine the data outputs for both regimes of observational measurements.

\subsection{Cosmology}

For a flat Friedmann–Lema\^{i}tre–Robertson–Walker (FLRW) background cosmology, we can take the metric to be described by
\begin{equation}
    ds^2 = dt^2 + a(t)^2 \left[ dr^2 + r^2\left(d\theta^2 + \sin^2\theta d\phi^2\right)\right]\,,
\end{equation}
where $a(t)$ is the scale factor, and which leads to the Ricci scalar
\begin{equation}
    R = 6 H^2 + 12 \dot{H}\,,
\end{equation}
where $H = \dot{a}/a$ is the Hubble parameter. Based on the equations of motion for the $f(R)$ model, Refs.~\cite{Sultana:2022qzn,Basilakos:2013nfa,Abuter:2018drb,Sultana:2022qzn,Rusyda:2022ywg} show that for small values of the parameter $b$, one is always able to find an analytic approximation to the Hubble parameter that works to a level of \textbf{precision to the level of $10^{-7}$ for small values of $\beta$.}
The model approximation gives a Friedmann equation
\begin{equation}
    \frac{H(z)^2}{H_0^2}= 1 - \Omega_m+(1+z)^3 \Omega _m + \beta \tilde{\alpha}(z) + \beta^2 \tilde{\beta}(z)\,,  \label{approx1}
\end{equation}
with the expansion of $H(z)$ defined in \cite{Abuter:2018drb,Sultana:2022qzn} from which the coefficients $\alpha$ and $\beta$ can be obtained. In the above expressions, we take $\Omega_{r} = 0$ to simplify the equations, since we use the late time data of the Universe, where the radiation density parameter is negligible. The corresponding expression fits for $\Lambda$CDM when $\beta\rightarrow 0$. The cosmological likelihood for the so defined EOS, $\chi^2_{cosmology}$, can be found in the Appendix.

\subsection{Orbital motion}

In order to solve the orbital motion one has to consider a general spherically symmetric metric \cite{misner1973gravitation}
\begin{equation}
    \label{eq:lineelement}
    ds^2=[1+\Phi(r)]dt^2-[1-\Phi(r)]dr^2-r^2d\Omega^2\,.
\end{equation}
where $\Phi(r)$ represents potentials and $d\Omega^2$ is the metric of a 2-sphere. For the low energy limit, the potential can be written as
\begin{equation}
    \label{eq:gravpot}
    \Phi(r)=-\frac{GM}{r} \frac{1 + \delta e^{-m_Y r}}{1 + \delta}\,,
    \end{equation}
where $M$ is the mass of the source of the gravitational field and $\delta$ and $\lambda$ are two parameters representing the strength and the scale length of the Yukawa-like modification of the gravitational potential \cite{Capozziello:2007ms}. Both parameters are also related to the $f(R)$ Lagrangian as \cite{Capozziello:2012ie,DeMartino:2018yqf,DeLaurentis:2018ahr,DeLaurentis:2018udw,Cardone:2011ze,Napolitano:2012fp,Capozziello:2008ny,Hees:2017aal,Zakharov:2018cbj,Capozziello:2020dvd,Capozziello:2007eu,Katsuragawa:2019uto}
\begin{align}
    \delta = f'_0 - 1, \qquad m_Y^2 = -\frac{f'_0}{6f''_0}\,,
\end{align}
where the derivatives of $f(R)$ are on a certain curvature $R_0$, where we assume that the S2 star experience on average the same curvature with small variance. Since we test the curvature of the S2 star we approximate it to be zero. Relativistic equations of motion for massive particles can be obtained from the geodesic equations for time-like geodesics of the metric in Eq.~\eqref{eq:lineelement} given by \cite{Benisty:2022txp}
\begin{equation}
    \label{eq:geodeiscequations}
    \frac{d^2x^\mu}{ds^2}+\Gamma^\mu_{\nu\rho}\frac{dx^\nu}{ds}\frac{dx^\rho}{ds} = 0\,.
\end{equation}
For this astrophysical system, the HS model gives parameters
\begin{equation}
    \delta = -2/\beta_{HS}, \quad m_{HS}^2 = \frac{(2 - \beta_{HS}) }{24} \beta_{HS}  \Lambda\,.
\end{equation}

For the Starobinsky model, we find the corresponding
\begin{equation}
\begin{split}
    \delta = -\frac{4 R_0}{\beta_{St} ^2 \Lambda  \left(\frac{R_0^2}{\beta_{St} ^2 \Lambda ^2}+1\right)^2}, \quad m_{St}^2 \approx \frac{\beta_{St} ^2 \Lambda }{384} (1 + \delta)\,.
\end{split}
\end{equation}
These simplified equations allow us to integrate the system by taking $\beta, \Lambda$ and $\delta$ (when applicable) as free parameters. From them, one can see that we can avoid using $\delta$ as a free parameter in the HS case, because it is directly connected to $\beta$. In the Starobinsky case, however, it is not possible to relate these two quantities since $\delta$ depends on the Ricci scalar $R_0$. For this reason for the HS case, we use as free parameters only $\beta$ and $\Lambda$, while for the Starobinsky case, we use $\beta, \Lambda$ and $\delta$. 

An additional consideration is the value of $\Lambda$, which is particularly important in the Starobinsky case due to its strong coupling to $\beta$. From Eqs.~(\ref{approx1}) and~(\ref{eq:geodeiscequations}) it is clear that $\Lambda \sim m^2$ and $m\sim 1/r$, so it has the units of length. The characteristic scale of the S2 system is that of its radius, so it is about 2000 AU. This means that the meaningful prior for $\Lambda$ is about $[0,10^{-3}]$. 

We start the first iteration using a sampling of the initial position $(x_0, y_0)$ and velocity $(\dot{x}_0, \dot{y}_0)$ of the corresponding star in the orbital plane at the epoch at $1995$. The true positions $(x_i, y_i)$ and velocities $(\dot{x}_i, \dot{y}_i)$ at all successive observed epochs are then calculated by numerical integration of equations of motion and projected into the corresponding positions $(x_i^c, y_i^c)$ in the observed plane (apparent orbit). There are three angles that we take into account: $\Omega$ is longitude of the ascending node, $\omega$ is longitude of pericenter and $i$ is the inclination. The transformation from the reference frame to our frame is via the rotation matrix $\left((l_1, l_2),(m_1, m_2)\right)$ where the expressions for $l_1, l_2, m_1$ and $m_2$ depend on three orbital elements:
\begin{equation}
\begin{array}{l}
    l_1=\cos\Omega\cos\omega-\sin\Omega\sin\omega\cos{i}\,, \\
    l_2=-\cos\Omega\sin\omega-\sin\Omega\cos\omega\cos{i}\,, \\
    m_1=\sin\Omega\cos\omega+\cos\Omega\sin\omega\cos{i}\,, \\
    m_2=-\sin\Omega\sin\omega+\cos\Omega\cos\omega\cos{i}\,. \\
\end{array}
\label{equ11}
\end{equation}
The S2 likelihood we use is:
\begin{equation}
 \chi_{S2}^2=((x_i-x_i^{obs})/\sigma_x)^2+(y_i-y_i^{obs})/\sigma_y)^2)^2,   
\end{equation}
where $x_i,y_i$ are the predicted orbital positions, $x_i^{obs},y_i^{obs}$ are the observed positions, and $\sigma_x$ and $\sigma_y$ are the respective errors in the positions. We take the angles as Gaussian priors since they are already well constrained in the literature. The only uniform priors that we run on the orbital motions are on the initial positions and velocities.

\subsection{Combined Constraint}

These two different systems have different energy scales. However, if the $f(R)$ is a fundamental model and not an approximate one, the additional parameter $\beta$ should be universal. Therefore,  we consider the combined likelihood $\chi^2$, as
\begin{equation}
    \chi^2_{Tot}=\chi_{cosmology}^2+\chi_{S2}^2\,,
\end{equation}
where the cosmological $\chi_{cosmology}^2$ include the expansion rate data and $\chi_{S2}^2$ includes the orbital data. The quantity we need to minimize is the separate or the combined $\chi^2$. For both datasets, the parameter $\beta$ appears directly and therefore we study the constraint from the partial and the combined systems on this parameter.

\begin{figure}[t!]
        \centering
\includegraphics[width=0.44\textwidth]{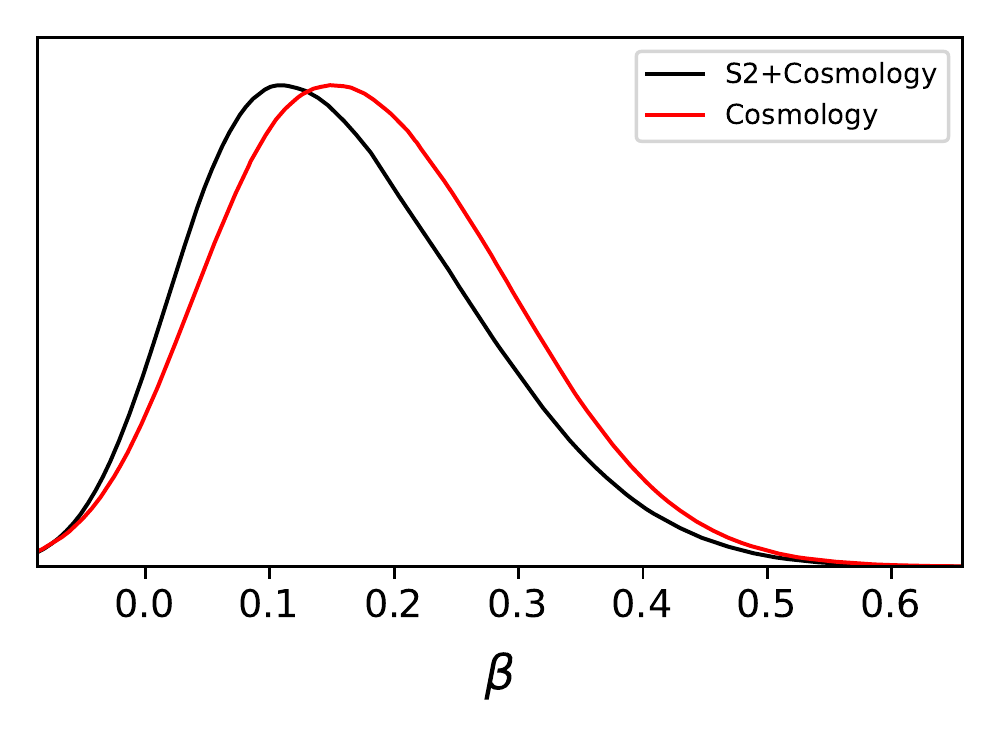}
\\
\includegraphics[width=0.44\textwidth]{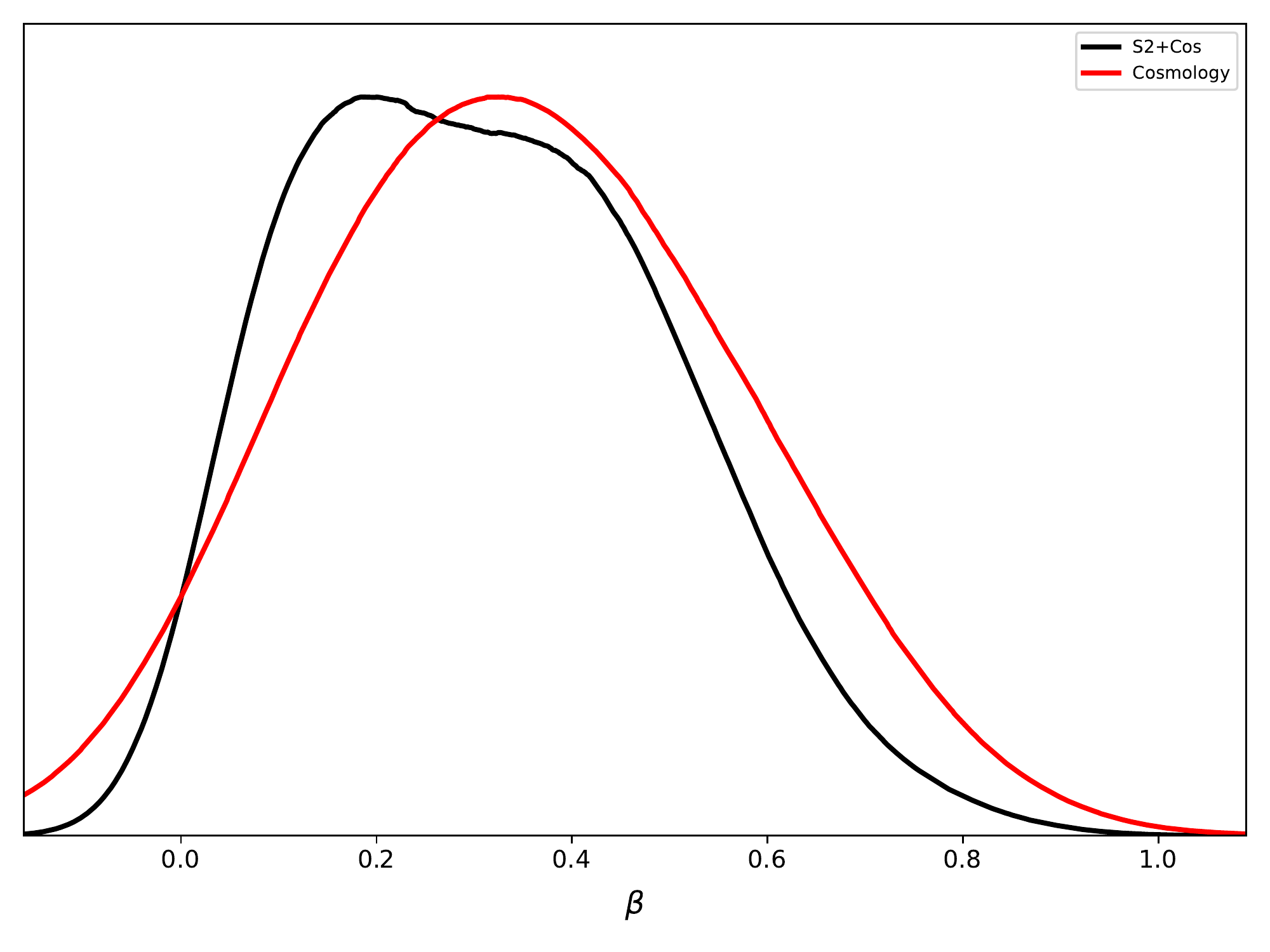}

\caption{\it{The posterior distributions for the additional parameter from the $f(R)$ gravity model: \textbf{Upper:} the Hu-Sawicki model, \textbf{Lower:} the Starobinski model. }} \label{fig:HS_Star_compare}   
\end{figure}

\section{Observational Data}
\label{sec:data}

In order to set constraints on the parameters of the model we shall consider various combinations of cosmological observations as well as data coming from the galactic center. The cosmological dataset include:

\begin{itemize}
\item \textbf{BAO} -- We use a combination of BAO points including various angular measurements and points from the most recent to date eBOSS data
release (DR16), which come as angular (DM) and radial (DH) measurements and their covariance. A description of the dataset can be found in Ref.~\cite{Staicova:2021ntm}. This choice of points allows us to integrate out the dependence on $H_0 r_d$ which allows us not to calibrate our cosmology with the early or the late Universe. The description of our  analytical marginalization approach can also be found in  \cite{Staicova:2021ntm}. 

\item \textbf{Pantheon} -- Type Ia Supernovae (SNeIa) distance moduli measurements from the Pantheon sample (SN) consisting of 1048 SNeIa in the range $0.01 < z < 2.3$ \cite{Scolnic:2017caz} divided into 40 bins. These measurements constrain the uncalibrated luminosity distance $H_0 d_L$, or in other words the slope of the late-time expansion rate. In this dataset, we marginalise analytically over $M_B$ and $H_0$. This is done again to avoid having these quantities as free parameters\footnote{During the work, a newer compilation was released called PantheonPlus which is available 
 \href{https://github.com/PantheonPlusSH0ES/DataRelease}{here}. We do not expect this new data set to appreciably change the results here}. 

\item \textbf{The orbital motion of the S2 star} - From the galactic center, our analysis uses publicly available astrometric and spectroscopic data that have been collected during the past thirty years. We use $145$ astrometric positions spanning a period from 1992.225 to 2016.38 from \cite{2017ApJ...837...30G}. The data come from speckle camera SHARP at the ESO New Technology Telescope \cite{1993Ap&SS.205....1H}, measurements were made using the Very Large Telescope (VLT) Adaptive Optics (AO). The data include the location and velocities but also the precession rate of the S2 star. We perform a Bayesian statistical analysis to constrain the additional parameter from the $f(R)$ gravity model.

\end{itemize}

We use an affine-invariant nested sampler \cite{ForemanMackey:2012ig} for the minimization of our likelihoods via the implementation of the open-source package $\text{Polychord}$ \cite{Handley:2015fda}. We also use the $\text{GetDist}$ package \cite{Lewis:2019xzd} for the analysis and illustration of our results. Based on ref.~\cite{Skilling:2006gxv},  the $\text{Polychord}$ estimates the evidence and the posterior simultaneously, by drawing $350$ (in our case) live points uniformly from the prior and updating them to find the peak of the posterior. We use modified precision criterion ($10^{-5}$) to increase the convergence.

For simplicity, we use the angles reported by the Gravity collaboration \cite{Abuter:2018drb, Amorim:2019hwp, Abuter:2020dou,} as a Gaussian prior. For the rest, we assume a uniform prior of the initial location and velocities. The priors on the parameters we use, thus are: $\beta\in[0,1]$, $x_0\in[1000,\,1500]$, $y_0\in[1300,\,1700]$, $v_{x0}\in[0,\,90]$, $v_{y0}\in[-150,\,-50]$, and $\Omega_m\in[0.1,0.5]$. Furthermore, we use the following Gaussian priors: $M_{Sgr*}=(4.261 \pm 0.012) 10^{6} M_{\odot}$ , $d=8.2467\pm 0.0093 \mathrm{kpc}$, $\Omega=228.171\pm 0.031$, $\omega=66.263\pm 0.031$, $i=134.567\pm 0.033$ The priors on  $\Lambda$ and $\delta$ are discussed in the next section.  

\begin{table}[t!]
    \centering
\begin{tabular}{|c|c|c|c|c|}
\hline
        System & $\beta$ & $\Lambda \,  (AU^{-2})$ & $\delta \left(10^{-3}\right)$ & $\Omega_m$\\
        \hline
         \multicolumn{5}{|c|}{HS}\\ 
         \hline
         C & $0.176\pm 0.11$  & - & - & $0.291\pm 0.014$  \\
         C+S2 & $0.154\pm 0.109$ &  $0.456\pm 0.299$ &  - & $0.292\pm 0.014$ \\
         \hline
         \multicolumn{5}{|c|}{Starobinsky}\\ 
          \hline
        C & $0.353\pm 0.219$  & - & - & $ 0.302\pm 0.012$  \\
         C+S2 & $0.309\pm 0.190$ & $0.459\pm 0.281 $ & $0.62\pm 57$ & $0.302\pm 0.011$\\
           \hline
\end{tabular}
    \caption{\it{The posterior distribution of different gravity theories with late time cosmology (noted as C) and the orbital motion of the S2 star (noted as S2 star).   }}
    \label{tab:posterior}
\end{table}

\section{Results}
\label{sec:res}
The results can be seen in Fig.~\ref{fig:HS_Star_compare}, which shows the posterior distribution for the parameter $\beta$ in both cases. The numerical values can be found in Table \ref{tab:posterior}. Due to the different parameters in the S2 case in the Hu-Sawicki and the Starobinsky model, we do not compare directly the S2 results. Instead, we present here the cosmology results, which are comparable since both depend only on the parameter $\beta$ and the "S2+Cosmology" results, which depend on $\beta$ and $\Lambda$ in the HS case, and on $\beta,\Lambda$ and $\delta$ in the Starobinsky case.

Numerically, the Starobinsky case presents a problem, since for big $\Lambda$ or $\delta$, the integration hits a numerical singularity. For this reason, we need to set either $\Lambda$ or $\delta$ small. The advantage of having $\Lambda$ small is that it corresponds to our expectations that it will be proportional to the characteristic units of length of the system. The disadvantage is that since it is coupled to $\beta$ in the S2 case, choosing a small prior for it will force $\beta$ to be less constrained. Since the posterior in this case do not improve on the cosmological result, we do not show it here. 

On the other hand, choosing a small $\delta$ does not bring such problems. Moreover, the parameter $\delta$ is expected to be small in order to recover GR. For this reason, on  Fig.~\ref{fig:HS_Star_compare}, we use as priors $\beta\in[0,1]$, $\Lambda\in[0,1]$ and $\delta\in[-0.1,0,1]$.  In this case, the MCMC is able to better constrain on $\beta$ in both cases and we see that for them, the addition of the S2 data improves the posterior on $\beta$ making it closer to GR. Note that in the Cosmology case, the prior on $\beta$ is rather large ( $\beta\in(0,10)$), yet the MCMC is able to constrain it very well from the data, well smaller than the prior interval. For the S2 and the combined case, such a large prior is not possible due to the mentioned singularity in the integration. The parameters $\Lambda$ and $\delta$ are not well constrained, because they cannot be fit solely based on the S2 equations and data, and they do not enter the associated cosmological models.  

One can compare these results with the ones presented in Ref.~\cite{Basilakos:2013nfa}. There, the model $f_1$CDM corresponds to our HS, and the model $f_2$CDM corresponds to Starobinsky. We see from Table III that indeed the Starobinsky model constraints come with a much larger error: $0.111 \pm 0.140$ for $f_1$CDM compared to $0.292 \pm 0.647$ for $f_2$CDM. We also observe this in our results: for cosmology we have $0.176 \pm 0.11$ vs $0.353 \pm 0.219 $ for Hu-Sawicki and Starobinsky respectively, and in the combined case we have $0.154 \pm 0.109$ vs  $0.309 \pm 0.191$. Note that in that article, they use numerous astrophysical datasets: SN, CMB, BAO and the growth rate data provided by the various galaxy surveys. In our work, we use only the marginalized BAO and SN datasets, thus some precision may be lost due to the lack of priors on $H_0$ and $r_d$. We see, however, that the matter density is very well constrained, as expected from the marginalized approach.  

The other parameters $\Lambda$ and $\delta$ as mentioned cannot be constrained efficiently from this approach. We study the effect of different choices of prior in the $S2$ system where the numerical singularities are more manageable. If one keeps the other priors fixed and changes just the prior on $\Lambda$ we see that a smaller prior leads to worse constrained posterior for $\beta$ in both models. This is due to $\Lambda$ being coupled to $\beta$ in both models and thus making it smaller immediately affects $\beta$. With respect to $\delta$, we find that if we keep the other parameters priors fixed, a decrease in the prior on $\delta$ leads to mild increase in the error of $\beta$. In this sense its effect on the system is much milder. The priors that we tested vary between $\Lambda\in[0,1]$ and $\Lambda\in[0,0.001]$ and $\delta\in[0,1]$ and $\delta\in[0,0.001]$. We find also that the sign of $\delta$ do not change the posterior on $\beta$.

\section{Discussion}
\label{sec:sum}

In this paper we have suggested a new combined approach, in which $f(R)$ gravity models screen different potentials that are tested directly in the galactic center. We have presented the bounds on $\beta$ from our combined astrophysics and cosmology system. We do not discuss $f'(R)$ and $f''(R)$ because in both systems, it depends on different quantities - for the cosmology it depends on $\beta$ while in S2, it depends on $\beta$ and $\delta$.

The discrepancy between the values of the Hubble constant $H_0$, the current expansion rate of the Universe, inferred from early-Universe measurements such as Planck CMB data \cite{Planck:2018vyg} and late-Universe measurements, such as the SH0ES collaboration \cite{Pan-STARRS1:2017jku}, has reached $5\sigma$ confidence. It was also suggested that the most promising method to accomplish this goal is by introducing new physics ~\cite{Abdalla:2022yfr}. $f(R)$ gravity models change with redshift and can lead to an $H_0$ estimate from CMB larger than that obtained from late-time probes. To solve the $H_0$ tension instead of modifying the matter content, the gravitational sector is modified in a manner that current cosmic dynamics is derived. $f(R)$ gravity could be a candidate for that. However, as we have shown in this work, a serious confrontation to the problem of cosmic tensions through $f(R)$ gravity must also incorporate astrophysical phenomenology, that is, any model must not only address the issue of tensions in cosmology but also retain the well-behaved evolution of stronger field systems such as S-type star orbits.

Many models within $f(R)$ gravity theories satisfy solar systems constraints whereas $f(R)$ models which evade the solar system constraints are equipped with a chameleon screening mechanism

~\cite{Brax:2008hh,Capozziello:2007eu,Katsuragawa:2019uto}. In this paper we suggest a novel method also to test these theories in comparison to $\Lambda$CDM, by adding the combined $\chi^2$ of both astrophysical and cosmological phenomenology.

Constraints on $f(R)$ gravity can be found in the literature \cite{Brax:2008hh}. In the notations we use, they vary from $\log_{10} \left[f'(R_0) - 1\right] < -4.79$ in galaxy clusters \cite{Cataneo:2014kaa}, to $\log_{10}\left[ f'(R_0)- 1\right] < -3$ from  GW 170817 \cite{Jana:2018djs} and $f'(R_0) - 1 < 3.7 \cdot 10^{-6}$ from the CMB \cite{Boubekeur:2014uaa} and $10^{-7} < f'(R_0) <10^{-4}$ from the fast predictions of the non-linear matter power spectrum ~\cite{Saez-Casares:2023olw}. While we cannot impose the strongest constraints on $f'(R)$ or even on $\beta$ due to the fact we choose a marginalised likelihood for cosmology, for the first time we find the constraints on the universal parameter $\beta$ from combined astronomical and cosmological dynamics. Since we compare two different systems, the curvature is different and the translation from the universal parameter $\beta$ to $f'(R_0)$ is not trivial. From the lower bound on $\beta$ we obtain that $GR + \Lambda$ is recovered. Therefore, our combined approach could be useful to test alternative theories in a novel way. It will be interesting in the future to extend tests in the strong-field regime to include black hole shadows, as was done in~\cite{Vagnozzi:2022moj} in a wide range of theories of modified gravity.

\acknowledgments 
We thank Prof. Salvatore Capozziello (Universita di Napoli ``Federico II'') for useful discussions and suggestions. D.B gratefully acknowledges the support of the Blavatnik and the Rothschild fellowships. D.S. is thankful to Bulgarian National Science Fund for support via research grants KP-06-N58/5. The authors would like to acknowledge funding from ``The Malta Council for Science and Technology'' through the ``FUSION R\&I: Research Excellence Programme''. This research has been carried out using computational facilities procured through the European Regional Development Fund, Project No. ERDF-080 "A supercomputing laboratory for the University of Malta". This paper is based upon work from COST Action CA21136 {\it Addressing observational tensions in cosmology with systematics and fundamental physics} (CosmoVerse) supported by COST (European Cooperation in Science and Technology).

\bibliographystyle{apsrev4-1}
\bibliography{ref}

\appendix
\section{Review on the marginalization process}

 {The cosmological measurements we use are outlined in \cite{Staicova:2021ntm}. The BAO measurements have two projections: the radial projection $D_H(z)= c/H(z)$ given by:}
\begin{equation}
\frac{D_H}{r_d} = \frac{c}{H_0 r_d} \frac{1}{E(z)},
\end{equation}
 and the tangential projection:
\begin{equation}
\frac{D_A}{r_d} = \frac{c}{H_0 r_d} f(z),
\end{equation}
where:
\begin{equation}
f\left(z\right) = \frac{1}{(1+z)  \sqrt{|\Omega_{K}|}  } \textrm{sinn}\left[|\Omega_{K}|^{1/2}\Gamma(z)\right]. 
\end{equation}
and $\textrm{sinn}(x) \equiv \textrm{sin}(x)$, $x$, $\textrm{sinh}(x)$ for $\Omega_{K}<0$, $\Omega_{K}=0$, $\Omega_{K}>0$ respectively and $\Gamma(z) = \int \frac{dz'}{E(z')}$.  The the angular diameter distance, $D_\textrm{A}$, is related to the comoving angular diameter distance trough $D_M= D_A (1+z)$.

The SNIa measurements are described by the luminosity distance $d_L(z)$ (related to $D_A$ by $D_A=d_L(z)/(1+z)^2$) and its distance modulus $\mu(z)$ through:
\begin{equation}
        \mu_B (z) - M_B = 5 \log_{10} \left[ d_L(z)\right] + 25  \,,
\label{eq:dist_mod_def}
\end{equation}
where $d_L$ is measured in units of Mpc, and $M_B$ is the absolute magnitude. 

{The $\chi^2$ for a DE model can be defined as:}
\begin{equation}
\begin{split}
\chi^2 = \sum_{i} \left[\vec{v}_{obs} - \vec{v}_{model}\right]^{T} C_{ij}^{-1} \left[\vec{v}_{obs} -  \vec{v}_{model} \right], 
\end{split}
\end{equation}
where $\vec{v}_{obs}$ is a vector of the observed points  {at each $z$ (i.e., $D_M/r_d$, $D_H/r_d$, $D_A/r_d$)}, $\vec{v}_{model}$ is the theoretical prediction of the model and $C_{ij}$ is the covariance matrix. For uncorrelated points {the covariance matrix is a diagonal matrix, and its elements are the inverse errors $\sigma_i^{-2}$. 

For BAO, it is possible to rewrite the vector as the dimensionless function $f(z)$ multiplied by the $\frac{c}{H_0 r_d}$ parameter and thus to eliminate the dependence of the result on $H_0, r_d $.  Following the approach in \cite{Lazkoz:2005sp,Basilakos:2016nyg,Anagnostopoulos:2017iao,Camarena:2021jlr},  we integrate over $H_0 r_d$ to get the final form of the marginalized $\chi^2$:
\begin{equation}
\tilde{\chi}^2 = C-\frac{B^2}{A} + \log\left(\frac{A}{2 \pi}\right).
\label{eq:chi2BAO}
\end{equation} 
where:
\begin{subequations}
\begin{equation}
A  = f^j(z_i) C_{ij} f^i(z_i),
\end{equation}
\begin{equation}
B = \frac{f^j(z_i) C_{ij} v_{model}^i(z_i) + v_{model}^j(z_i) C_{ij} f^i(z_i)}{2},
\end{equation}
\begin{equation}
C = v_j^{model} C_{ij} v_i^{model}.
\end{equation}
\label{eq:termChi2}
\end{subequations}

For the Supernova data, following the approach in (\cite{DiPietro:2002cz,Nesseris:2004wj,Perivolaropoulos:2004yr,Lazkoz:2005sp}), we marginalize over $M_B$ and $H_0$, so that the integrated $\chi^2$ becomes:
\begin{equation}
\tilde{\chi}^2_{SN} = D-\frac{E^2}{F} + \ln\frac{F}{2\pi},
\end{equation}
where:

\begin{align}
D = \sum_i \left( \Delta\mu \, C^{-1}_{cov} \, \Delta\mu^T \right)^2,
 \nonumber \\
E = \sum_i \left( \Delta\mu \, C^{-1}_{cov} \, E \right),\nonumber \\
F = \sum_i  C^{-1}_{cov}.
\end{align}
where $\Delta\mu =\mu_{}^{i} - 5 \log_{10}\left[d_L(z_i)\right)$, $E$ is the unit matrix, and $C^{-1}_{cov}$ is the inverse covariance matrix of the dataset. Here $\mu_{}^{i}$ is the observed luminosity, $\sigma_i$ is its error.  The total covariance matrix is given by $C_{cov}=D_{stat}+C_{sys}$, where $D_{stat}=\sigma_i^2$ comes from the measurement and $C_{sys}$ is provided separately \cite{Deng:2018jrp}.

There is a difference between $\tilde{\chi}^2_{BAO}$ and $\tilde{\chi}^2_{SN}$ because for the BAO we removed the dependence of $c/{H_0 r_d}$, which is multiplied to the $f(z)$, while for SN, the parameter, $\bar{M_B}$ is added to the value of $\mu$.

Thus the combined likelihood for cosmology becomes: 
\begin{equation}
\tilde{\chi}^{2}_{cosmology}= \tilde{\chi}_{BAO}^{2} + \tilde{\chi}_{SN}^{2}.
\end{equation}

For it, the values of $H_0r_d$ and $M_B$ and $H_0$ for the BAO and the SN respectively don't change the marginalized $\tilde{\chi}^{2}_{cosmology}$.

\end{document}